\documentclass{article}

\usepackage{arxiv}

\usepackage[utf8]{inputenc} % allow utf-8 input
\usepackage[T1]{fontenc}    % use 8-bit T1 fonts
\usepackage{hyperref}       % hyperlinks
\usepackage{url}            % simple URL typesetting
\usepackage{amsfonts}       % blackboard math symbols
\usepackage{nicefrac}       % compact symbols for 1/2, etc.
\usepackage{microtype}      % microtypography
\usepackage{graphicx}
\usepackage{natbib}
\usepackage{doi}
\usepackage{amsmath}
\usepackage{natbib}         
\usepackage{booktabs}        
\usepackage{enumitem}
\usepackage{multirow}
\usepackage{threeparttable}
 \usepackage[normalem]{ulem}
 \useunder{\uline}{\ul}{}

\title{Identifying Treatment Effect Heterogeneity with Bayesian Hierarchical Adjustable Random Partition (BHARP) Model in Adaptive Enrichment Trials}

\author{ 
\textbf{Xianglin Zhao}\\
Department of Epidemiology, Biostatistics, and Occupational Health\\
McGill University, Montreal, Quebec, Canada\\
\texttt{xianglin.zhao@mail.mcgill.ca}\\
\And
\textbf{Shirin Golchi}\\
Department of Epidemiology, Biostatistics, and Occupational Health\\
McGill University, Montreal, Quebec, Canada\\
\And
\textbf{Jean-Philippe Gouin}\\
Department of Psychology\\
Concordia University, Montreal, Quebec, Canada
\And
\textbf{Kaberi Dasgupta}\\
Department of Medicine\\
McGill University, Montreal, Quebec, Canada
}

\date{December 2025\\Preprint.}

\renewcommand{\shorttitle}{Identifying Treatment Effect Heterogeneity with BHARP Model in Adaptive Enrichment Trials}

\hypersetup{
  pdftitle={Identifying Treatment Effect Heterogeneity with Bayesian Hierarchical Adjustable Random Partition in Adaptive Enrichment Trials},
  pdfauthor={Xianglin Zhao, Shirin Golchi, Jean-Philippe Gouin, Kaberi Dasgupta},
  pdfsubject={Statistics - Applications (stat.AP)}, 
  pdfkeywords={Bayesian hierarchical model; dynamic information borrowing; finite mixture model; model uncertainty; precision medicine; reversible-jump Markov chain Monte Carlo.}
}

\begin{document}
\maketitle

\begin{abstract}
Motivated by an adaptive enrichment trial of physical activity interventions in dyads with type 2 diabetes, we develop a method to investigate treatment effect heterogeneity and estimate subgroup-specific responses.  While several information-borrowing methods have been proposed to address this problem, existing partitioning-based methods often select a single partition which results in under-representation of the uncertainty in the borrowing structure. 
 We propose the Bayesian Hierarchical Adjustable Random Partition (BHARP) model, a self-contained framework that applies a finite mixture model with an unknown number of components to explore the partitioning space accounting for model uncertainty. The BHARP model jointly estimates subgroup-specific effects and the latent heterogeneity patterns without requiring manual calibration. Posterior sampling is performed via a custom reversible-jump Markov chain Monte Carlo sampler tailored to partitioning-based information borrowing in clinical trials. Simulation studies across a range of treatment effect heterogeneity scenarios show that the BHARP model achieves competitive accuracy and precision while identifying the latent heterogeneity structure with high computational efficiency. In addition, we demonstrate the applicability of the model in the context of a multi-arm adaptive enrichment trial investigating physical activity interventions in patients with type 2 diabetes. 
\end{abstract}

% keywords can be removed

\section{Introduction}
\label{s:intro}

%Treatment effect heterogeneity (TEH) refers to systematic variation in treatment effects across prespecified subgroups. Understanding TEH is central to precision medicine as TEH structures can reveal underlying biological mechanisms and guide individualized clinical decision making. 
Motivated by the Partner Step T2D study, an adaptive enrichment trial evaluating behavioral interventions to increase step counts among dyads with type 2 diabetes, we consider the challenge of identifying treatment effect heterogeneity (TEH) across predefined subgroups.
In this trial, relationship quality and weight concordance are potential effect modifiers, and investigators aim to identify subgroups that benefit most and to determine which subgroups exhibit similar responses. 
However, identifying TEH patterns and estimating subgroup-specific responses within a single clinical trial remains challenging \citep{freidlin2010design, rosenblum2017enrich, xuzhuLee2020, thall2021enrich}, which motivates information borrowing across subgroups. 
% In practice, however, TEH presents two fundamental challenges: (i) estimating subgroup‐specific responses and (ii) identifying the structure that characterizes similarity across subgroupsAddressing both challenges simultaneously requires statistical methods that can borrow information adaptively while identifying heterogeneity structure.

Bayesian hierarchical models (BHMs) \citep{thall2003, berry2013bhm} improve estimation efficiency by borrowing  across all subgroups, and perform well when subgroup responses are similar \citep{chenLee2020, zheng2022borrowing}. Yet under TEH, the assumption of full exchangeability induces excessive shrinkage, biasing subgroup‐specific estimates and masking the TEH of interest. Several extensions of BHM attempt to mitigate this limitation. Some calibrate the borrowing strength empirically \citep{cbhm2018}, while others relax full exchangeability assumption \citep{JiangYuan2021, kang2021}. For example, \citet{exnex2016} combine a pooled distribution with subgroup‐specific components, and \citet{chenLee2019BACIS} classify subgroups into responsive versus nonresponsive clusters before fitting separate BHMs within each cluster. Although effective in reducing overshrinkage, these approaches obscure finer‐scale TEH patterns.

Another class of models introduce an explicit partition of subgroups, grouping them into clusters with similar responses \citep{zhouji2024review,luLee2023,lin2021}. Within each cluster, parameters are treated as exchangeable to enable borrowing, while parameters across clusters are assumed nonexchangeable to preserve heterogeneity. In most of these approaches, however, the partition is introduced primarily as a device to construct an information-borrowing structure rather than as a target for inference. Typically, a single ``optimal" partition is selected according to a predefined criterion, and subsequent inference is conducted conditional on this chosen partition \citep{geng2020, chenLee2020, lin2021, luLee2023}. In fact, different criteria can favor different partitions, and even under a fixed criterion multiple partitions often achieve similar scores, yet this model uncertainty is not propagated to the final subgroup‐level estimates.
In contrast, \citet{Psioda2021} perform Bayesian model averaging over partitions with closed‐form posteriors, thereby formally accounting for uncertainty in the information–borrowing structure. Nevertheless, the number of possible partitions grows exponentially with the number of subgroups (e.g., {115,975} for 10 subgroups), rendering exhaustive enumeration infeasible for trials with a moderate to large number of subgroups and motivating methods that can explore the partition space more efficiently.

% BHARP
%
To address these limitations, we propose the Bayesian Hierarchical Adjustable Random Partition (BHARP) model, a flexible and fully Bayesian framework that identifies TEH structures in parallel with estimating subgroup-specific treatment effects. The BHARP model is formulated as a finite mixture model (FMM) with an unknown number of components---a ``mixture of finite mixtures"---and treats the partition structure as a random quantity. By averaging over the partition space through posterior sampling, BHARP formally accounts for model uncertainty, directly captures subgroup similarity patterns, and automatically adjusts the degree of information borrowing. We implement a reversible-jump Markov chain Monte Carlo (rjMCMC) algorithm customized for clinical trial settings to traverse models with varying dimensionality, thereby jointly estimating subgroup-specific parameters and latent borrowing structures. Although, rjMCMC has been used in other contexts such as variable selection \citep{lin2021}, existing implementations do not accommodate partition-based information borrowing. 
%To our knowledge, our implementation represents the first use of rjMCMC for dynamic Bayesian partitioning of heterogeneous treatment effects.

We evaluate the proposed model through extensive simulation studies spanning a range of TEH scenarios. We compare the BHARP model with simplified parametric models and Bayesian additive regression trees, focusing on the performance of estimates and the identification of TEH patterns. We further embed the BHARP model within a Bayesian adaptive trial framework with subgroup enrichment and early stopping rules at interim analyses. Finally, we demonstrate the applicability of the model in the context of an adaptive enrichment trial for the Partner Step T2D study. Across simulations and the real-data application, BHARP reliably recovers prespecified TEH patterns and supports data-driven decision-making in adaptive trial settings.

The remainder of the paper is organized as follows. 
Section~\ref{s:methodology} introduces the BHARP framework. 
To compare the proposed method with alternative methods, Section~\ref{s:simulation} investigates estimation accuracy and precision in a simulation study. Section~\ref{s:application} presents simulations for a hypothetical design scenario inspired by a real-world multi-arm trial. Section~\ref{s:discussion} follows with a discussion.

\section{Methodology}
\label{s:methodology}

\subsection{Model}
\label{ss:model}

Consider a clinical trial comparing treatment response across $K$ pre-defined subgroups. Let $\boldsymbol Y_{k}$ denote the continuous outcomes observed in subgroup $k$, modeled as
$\boldsymbol Y_{k} \sim N(\theta_{k}, \varsigma^{-1}), k=1,2,\ldots,K. $
To simplify exposition, a common variance $\varsigma^{-1}$ is assumed across subgroups. This assumption may be relaxed to accommodate heteroskedasticity.
%This assumption is not essential, as the proposed framework can naturally accommodate heteroskedasticity. Because the method is intended for large-sample adaptive trials, the homoscedastic specification is adopted mainly for notational convenience rather than as a substantive restriction.     

To borrow information between subgroups with similar responses, we assume the mean responses $\boldsymbol\theta = (\theta_1, \ldots, \theta_K)$ arise from a FMM with an unknown number of mixture component $q \in \{1,\ldots,K\}$.  This induces a partitioning structure where subgroup-specific parameters from a common mixture component correspond to a cluster with similar responses. We introduce the vector $\boldsymbol{z} = (z_1,\ldots,z_K)$ to indicate the component allocation of the $K$ subgroups. Therefore, we have,
\[
\begin{aligned}
&\theta_{k}|\boldsymbol{\mu},\boldsymbol{\sigma},  z_k \sim  N(\mu_{z_{k}},{\sigma_{z_{k}}}),\\
&\boldsymbol{z}|q,\boldsymbol w  \sim Multinomial(q;\boldsymbol{w}).
\end{aligned}
\]
The priors are specified as follows:
\[
\begin{aligned}
\varsigma &\sim Gam(a_{cell}, b_{cell}), \quad
P(q) \propto q^\alpha, \quad
\boldsymbol{w}\mid q \sim Dirichlet(1,\ldots,1), \\
\tau &\sim Gam(a_{\text{between}}, b_{\text{between}}), 
\quad \mu_t \mid \tau \sim N(0,\tau^{-1}), \quad
\sigma_t \sim InvGam(a_{\text{within}}, b_{\text{within}}).
\end{aligned}
\]
where $Gam(a,b)$ and $InvGam(a,b)$ denote gamma and inverse-gamma distributions with shape parameter $a$ and rate parameter $b$, and $N(c,p^{-1})$ denotes a normal distribution with mean $c$ and precision $p$. 
While we describe the model for a single-arm trial for simplicity, it can be readily extended to $I$ intervention arms with the same mixture formulation applied to each arm. In this multi-arm setting, each combination of intervention $i$ and subgroup $k$ constitutes a ``cell", i.e., the basic unit at which treatment responses are modeled and borrowing occurs. Because the proposed method is highly automated and computationally efficient, it is particularly well-suited for platform or multi-arm trials aimed at exploring TEH.

Below we explain the rationale for the choices of prior distributions as well as the specification of hyperparameters. Overall, providing minimal but targeted prior information is essential to ensure that the information-borrowing mechanism aligns with TEH patterns of practical interest. Importantly, these priors do not introduce substantive bias; instead, they clarify the division of labor across model layers---preventing, for example, the within-component variance from absorbing heterogeneity that should be represented by the dynamic partition---thereby preserving clinical interpretability and relevance of the inferred TEH structure.

%\textbf{Precision $\varsigma$}
%is assigned a Gamma prior. The semi-conjugate form yields closed-form full conditional updates in the nested Gibbs steps within rjMCMC algorithm.  Compared to other parameters, $\varsigma$ is well-informed by the data, so a vague prior distribution work well. 

\textbf{Allocation vector $\boldsymbol z$} 
assigns each subgroup $k$ to one of the $q$ mixture components through $z_k=t$, indicating that the subgroup's response is governed by component $t$. Given the number of components $q$ and weights $\boldsymbol w$,  $\boldsymbol z$
follows a multinomial distribution. The vector $\boldsymbol z$ induce a partition  $\boldsymbol{\Omega}=\{\Omega_t:\Omega_t=\{k:z_k=t\}, \Omega_t \neq \emptyset\}$, where clusters are mutually exclusive and collectively exhaustive. Importantly, this partition is the primary inferential target: it determines which subgroups are considered practically equivalent, where the exchangeability assumption holds, and how information borrowing occurs. Although FMMs are non-identifiable up to permutation of component labels, this lack of inferential meaning poses no difficulty in our framework. All posterior quantities of interest---such as co-clustering probabilities and subgroup-level estimates---are invariant to label permutations. Furthermore, although reversible-jump moves change the dimensionality of the parameter space, each Markov chain Monte Carlo (MCMC) chain maintains internally coherent labeling across iterations within the chain, and inference remains unaffected without artificial constraints.

\textbf{Number of mixture components $q$} determines the maximum number of 
clusters the model can represent. Note that this is not the number of actually-formed clusters. Even when $q$ is large, the realized partition may place all subgroups in a single cluster. Thus, $q$ should be interpreted as ``model capacity", while the effective clustering is governed by $z$. We assign a prior of the form $P(q) \propto q^\alpha$, where larger $\alpha$ favor greater flexibility and conservative borrowing %, despite being somewhat counterintuitive. 
We recommend $\alpha=2$ to prevent overly coarse clusters. Importantly, posterior inference of $q$ remains data-driven, and the model capacity is adjusted to the underlying TEH structure: when the true structure contains a single cluster, the posterior of $q$ concentrates at 1; when multiple clusters exist, larger values of $q$ are supported under the posterior (See Section~\ref{s:simulation}).

\textbf{Component weights $\boldsymbol w$}
are assigned a symmetric Dirichlet prior $Dirichlet(1,\ldots,1)$ which is weakly informative and does not favor any particular allocation pattern. If prior knowledge suggests structural differences among certain subgroups, asymmetric priors or identifying constraints may be incorporated.

\textbf{Between-component precision $\tau$} determines the overall scale on which component means $\boldsymbol \mu$ may vary. In our model, this ``between-component" layer keeps the the components from drifting away from plausible values but induces little information-borrowing. Most of the information borrowing is instead driven by the partition layer. We adopt a weakly informative gamma prior with shape $a_{\text{between}}$ and rate $b_{\text{between}}$, constraining $\tau$ to a reasonable magnitude on the effect-size scale without imposing strong shrinkage. When the true number of clusters is small, a noninformative prior can yield implausible component locations and allocations.

\textbf{Within-component variances $\boldsymbol \sigma$}
serve primarily a computational role by preventing discontinuous likelihood. The partition structure determines information borrowing, and its uncertainty is captured by FMM with an unknown number of components. Although $\boldsymbol \sigma$ does account for some uncertainty and adaptively modulate the borrowing strength within clusters, these are secondary consequences. Primarily, $\boldsymbol \sigma$ provides a small amount of continuous variability around component center, ensuring continuity and stable reversible-jump transitions. To maintain the interpretation of co-clustering as practically equivalent subgroup effects, we assign an informative inverse-gamma prior that concentrates mass on values corresponding to negligible heterogeneity on the effect-size scale. This prevents the model from implicitly redefining what constitutes meaningful TEH and ensures that subgroups with clinically meaningful differences are assigned to different clusters. The prior should be concentrated enough to promote cohesion among co-clustering subgroups, yet sufficiently diffuse to preserve the continuity required for efficient and well-behaved reversible-jump updates.

\subsection{Systematic elicitation guideline}
If substantive prior knowledge about plausible TEH or component-level parameters is available, it should be incorporated into the prior specification. In settings where such information is limited, we provide a general guideline for eliciting hyperparameters under minimal prior assumptions.
Generally, to interpret between-component differences as clinically meaningful, the component means $\boldsymbol \mu$ should be well-separated relative to within-component variances $\boldsymbol{\sigma}$. 
For a pair of subgroups whose difference in effect sizes exceeds a clinically negligible range, they should, a priori, be more likely to be assigned to different components than to the same one.
This principle guides the selection of $a_{\text{within}}$, $b_{\text{within}}$, $a_{\text{between}}$ and $b_{\text{between}}$. We compute the crossover threshold $\delta$, defined as the largest value such that the prior density of within-component differences exceeds that of differences between component means over the interval $(-\delta,\delta)$. This threshold reflects the minimal heterogeneity level at which two subgroups are more likely to be assigned to different components rather than the same one. By verifying that $\delta$ does not exceed the clinically significant difference we aim to detect, we ensure that the prior specification supports practically relevant partitions. For example, suppose a difference of 0.4 in effect sizes is regarded as clinically meaningful. One can then specify hyperparameters such that the resulting crossover value satisfies $\delta < 0.4$, so that subgroups differing by 0.4 are a priori more likely to be allocated to different components than to the same one; as a result, the induced partitioning reflects the TEH structure of interest. %In our simulation study (Section~\ref{ss:hyperparam}), the specified hyperparameters yield $\delta \approx 0.31$, which lies below the gap between heterogeneous response levels of interest and is therefore consistent with this guideline.
%We note that \citet{nobilegreen2000} proposed to specify hyperparameters through theoretical calculation in the absence of prior information. This approach is less suited to clinical trials, where expert knowledge on plausible TEH and range of effect sizes should be utilized. 

%
\subsection{Computation}
\label{ss:computation}

We employ a rjMCMC algorithm \citep{rjMCMC}, an extension of MCMC that enables transitions across parameter spaces with varying dimensionality. The sampler jointly updates the partition structure and model parameters, thereby capturing uncertainty in both.
Following the split–merge framework of \citet{nobilegreen2000}, each iteration consists of two steps: (i) a within-model update, where all parameters are drawn from their full conditional distributions via Gibbs sampling, and (ii) a between-model update, where a transdimensional proposal is generated and accepted with a probability that maintains detailed balance.

At the transdimensional step, the algorithm randomly proposes either a ``merge" or a ``split". The split and merge proposals are mutually reversible: each split has a unique inverse merge, and vice versa. The parameters $w$, $\mu$ and $\sigma$ of the proposed component(s) are constructed deterministically from those of the current component(s). To maintain continuous indexing, in a merge move the last component is absorbed into another randomly selected component; in a split move, a randomly chosen component gives rise to a new one appended to the end of the list. Component indices do not encode any ordinal information and play no inferential role—they merely provide a consistent labeling scheme within each MCMC chain. The acceptance probability depends on the likelihood and prior ratios, the forward–reverse proposal probabilities, and the Jacobian of the transformation. Accepted proposals move the chain across model dimensions; otherwise, the chain remains at its current state.

%Implementation
The algorithm is implemented in C++ via \texttt{Rcpp}, enabling efficient sampling with dynamically varying parameter dimensions. An R wrapper function provides a user-friendly interface. Additional implementation details are provided in the Supplementary Material, S1.

\subsection{Adaptive Design}
\label{ss:design}

In this section, we describe an adaptive enrichment trial design that modifies the enrollment criteria to focus on responsive subgroups and allows for early termination of treatment arms once sufficient evidence is accumulated. Decisions are based on the posterior distribution of $\theta_{ik}$, i.e., the expected response of subgroup $k$ to intervention $i$, conditional on data $\mathcal{D}_\ell$ up to analysis $\ell=1,2,\ldots,\mathcal{L}$, where $\mathcal{L}$ denotes the final analysis. 
The decisions rely on clinically meaningful boundaries for efficacy and futility ($x_E$ and $x_F$, respectively) along with posterior probability decision thresholds ($\mathcal{P}_{\ell, E}$ and $\mathcal{P}_{\ell, F}$). These thresholds may be set more stringently at early analyses to mitigate the impact of early-stage sampling variability.

At each interim analysis $\ell=1,\ldots,\mathcal L-1$ the following sequential procedures are applied: 
\textbf{Enrichment:} Conclude futility and deactivate recruitment of subgroup $k$ in arm $i$ if $P(\theta_{ik}\leq x_{F} | \mathcal{D}_{\ell})>P_{\ell, F}$; 
\textbf{Efficacy:} For interventions remaining active in at least one subgroup, test the efficacy hypothesis in each active cell $(i,k)$. Conclude efficacy if $P(\theta_{ik} > x_{E} | \mathcal{D}_{\ell})>P_{\ell, E}$;  
\textbf{Arm termination:} Cease randomization to intervention $i$ if all associated subgroups have reached any conclusion.
\textbf{Continued Accrual:} If the trial continues beyond interim analysis $\ell$, recruitment continues for all unresolved subgroup--arm combinations until the next analysis $\ell + 1$. At the final analysis $\ell=\mathcal{L}$,  for each unresolved subgroup--arm combinations, conclude futility if $P(\theta_{ik} \leq x_F|\mathcal{D}_\mathcal{L})>\mathcal{P}_{\mathcal L ,F}$ and conclude efficacy if $P(\theta_{ik}>x_E|\mathcal{D}_\mathcal{L})>\mathcal{P}_{\mathcal L ,E}$.

The adaptive decision framework is guided by three primary considerations: early identification and deactivation of futile subgroups enables the enrichment of promising subgroups in later trial stages; accrual continues in subgroups demonstrating efficacy until sufficient evidence is collected for the entire intervention arm, as premature discontinuation of access to potentially beneficial treatments during ongoing evaluation raises ethical concerns; and the design allocates resources across arms with different TEH complexity. Interventions with simpler TEH structures (e.g., homogeneity or readily identifiable heterogeneity) benefit more from dynamic information borrowing and reach conclusions faster. This enables early termination of such arms and reallocation of sample size to arms with more complex TEH structures that require additional data to identify.

\section{Simulation Study}
\label{s:simulation}

In this section, we compare the BHARP model with parametric models with simplified structure as well as a flexible non-parametric model in simulation studies assessing the accuracy and precision of point estimates in a non-comparative analysis across diverse TEH scenarios.

\subsection{Comparator Methods}
\label{ss:comparing}

To assess the contribution of each element in the proposed BHARP framework, we compare it with the following alternative methods: a Bayesian independent (IND) model, a basic BHM \citep{berry2013bhm} and a FMM with fixed number of components (BLAST) \citep{ChuYuan2018blast}. Each comparator is derived by excluding specific elements from the BHARP model, thereby highlighting the contribution of each element. 
Additionally, to benchmark against a flexible yet less interpretable non-parametric method, we include Bayesian additive regression trees (BART) \citep{2010bart}. 
Detailed model specifications are given below.

\textbf{IND:} This model treats each subgroup–arm cell as an independent sub-trial and serves as a no-borrowing benchmark.
For subgroup $k$, the outcome model is identical to that in BHARP, but the cell-specific parameter is assigned an independent weakly informative prior $\theta_k \sim N(0, 10)$. This prior is sufficiently diffuse on the standardized effect scale and does not induce any cross-subgroup pooling, so the posterior distribution of each $\theta_k$ is updated separately.

\textbf{BHM:}  A standard BHM is applied within each intervention arm, assuming full exchangeability across subgroups. The subgroup-specific parameters follow $\theta_{k} \sim N(0,\tau_{BHM}^{-1})$, with prior for $\tau_{BHM}$ matched to that of $\tau$ in the BHARP model. Conceptually, this model excludes the partitioning FMM layer from the BHARP model. 
While both the BHM and the BHARP model include this weak-borrowing layer, a distinction lies in its role: the BHM pools all subgroups and relies on this layer as the sole borrowing mechanism, whereas in the BHARP model, it regulates the component centers and borrows primarily through the partition structure. %Furthermore, the purpose of the weak-borrowing layer is different: it serves as the primary mechanism of information sharing in the BHM, but contributes only marginally in the BHARP model, where primary borrowing occurs through the partition structure.

\textbf{BLAST:} This framework also contains FMM-based information borrowing. The only difference is it assumes that the component number $q$ to be fixed. Rather than assigning a prior to $q$, its value is selected from ${1,2,3}$ based on the deviance information criterion (DIC). Comparing with the BLAST method allows us to investigate the added flexibility in the BHARP model by unknown number of components. To make the results comparable, we modify the original BLAST model to have identical hyperparameters and likelihood as the BHARP model. 

\textbf{BART:} This method uses sum of shallow regression trees to learn the relationship between subgroup indicators and response. %In BART, the subgroup indicators are treated as categorical covariates. 

In the simulation study, methods are compared over 500 simulated data sets under each scenario. For each data set, inference for the parametric models is made using four MCMC chains with 2,000 iterations each. The posterior median is used as the point estimator.  BART is implemented with a single chain with 8,000 iterations. %due to the lack of built-in multi-chain support.

\subsection{Hyperparameter Specification}
\label{ss:hyperparam}

In this subsection, we specify hyperparameters $a_{\text{cell}}$, $b_{\text{cell}}$, $a_{\text{between}}$, $b_{\text{between}}$, $a_{\text{within}}$ and $b_{\text{within}}$ in the simulation study. All priors are defined on a standardized effect-size scale, which facilitates generalizability of the prior choices to other settings. The prior for cell-level precision $\varsigma$ is specified as $Gam(a_{\text{cell}}, b_{\text{cell}})=Gam(5, 6)$,  allowing sufficient flexibility for unit variance.  Compared to other parameters, $\varsigma$ is well-informed by the data, so a vague prior distribution works well. 
For the precision of component means, we specify a moderately informative prior $Gam(a_{\text{between}},b_{\text{between}})=Gam(4, 4)$,  sufficiently vague to accommodate a wide range of plausible effect sizes. This prior allows clinically meaningful heterogeneity while avoiding excessive shrinkage. Sensitivity analyses show that $\tau$ is robust to prior specifications (see Supplementary Material, S4).

We fix the mode of standard deviation of components at 0.1, consistent with the clinically negligible difference in effect sizes. To calibrate the informativeness of the within-component prior relative to subgroup sample size, we set $a_{\text{within}}$ to be twice as large as the subgroup size. In our simulation settings the smallest subgroup size is 35, yielding $a_{\text{within}}=70$ and $b_{\text{within}}=0.71$. The resulting crossover value is approximately 0.31 which remains below the gap between heterogeneous response levels of interest. This alignment with the TEH assumption ensures the prior supports meaningful and clinically interpretable partitions. 

\subsection{Simulation Scenarios}
\label{ss:simulation_scenarios}

We constructed a set of twelve scenarios (Table~\ref{t:scenarios}) that span a broad spectrum of TEH patterns that may be encountered in clinical trials, including no heterogeneity, sparse strong signals, moderate and unbalanced cluster structures, within‐cluster variability, small rare subgroups, and multi‐cluster configurations. 
This design allows us to evaluate both the accuracy of effect estimation and the ability to recover the underlying partition structure under increasingly challenging settings.
\begin{table}[t]
\caption{True subgroup effects ($\theta_k$) and subgroup sample-size configurations for the twelve simulation scenarios. Each scenario specifies ten subgroup-specific means, and outcomes for subgroup $k$ are generated from $N(\theta_k,1)$. The rightmost column reports the per-subgroup sample sizes. Daggers denote subgroups with a small sample size (20 observations).}
\label{t:scenarios}
\centering
\setlength{\tabcolsep}{4.25pt}
\resizebox{\textwidth}{!}{%
\begin{tabular}{lccccccccccc}
\toprule
& \multicolumn{10}{c}{True subgroup effects ($\theta_k$)} & Subgroup \\
\cmidrule(lr){2-11}
Scenario &
$\theta_1$ & $\theta_2$ & $\theta_3$ & $\theta_4$ & $\theta_5$ &
$\theta_6$ & $\theta_7$ & $\theta_8$ & $\theta_9$ & $\theta_{10}$ &
size \\
\midrule
A & 0 & 0 & 0 & 0 & 0 & 0 & 0 & 0 & 0 & 0 & 35 \\
B & -0.08 & -0.08 & -0.04 & -0.04 & 0 & 0 & 0.04 & 0.04 & 0.08 & 0.08 & 35 \\
C & 0 & 0 & 0 & 0 & 0 & 0 & 0 & 1.3 & 1.3 & 1.3 & 35 \\
D & -0.08 & -0.04 & -0.04 & 0 & 0.04 & 0.04 & 0.08 & 1.22 & 1.26 & 1.3 & 35 \\
E & 0 & 0 & 0 & 0 & 0 & 0 & 0 & 0.65 & 0.65 & 0.65 & 70 \\
F & -0.08 & -0.04 & -0.04 & 0 & 0.04 & 0.04 & 0.08 & 0.61 & 0.65 & 0.69 & 70 \\
G & -0.08 & -0.04 & -0.04 & 0 & 0.04 & 0.04 & 0.08$^{\dagger}$ & 0.61$^{\dagger}$ & 0.65 & 0.69 & 70,20 \\
H & 0 & 0 & 0 & 0 & 0 & 0.65 & 0.65 & 0.65 & 0.65 & 0.65 & 70 \\
I & -0.08 & -0.04 & 0 & 0.04 & 0.08 & 0.57 & 0.61 & 0.65 & 0.69 & 0.73 & 70 \\
J & -0.08 & -0.04 & 0 & 0.04 & 0.08$^{\dagger}$ & 0.57$^{\dagger}$ & 0.61 & 0.65 & 0.69 & 0.73 & 70,20 \\
K & 0 & 0 & 0 & 0 & 0.65 & 0.65 & 0.65 & 1.3 & 1.3 & 1.3 & 70 \\
L & 0 & 0 & 0 & 0 & 0 & 0 & 0 & 0.65 & 0.65 & 1.3 & 70 \\
\bottomrule
\end{tabular}%
}
\end{table}

Scenario A contains no TEH, whereas Scenario C features strong but sparse signals, where only a few subgroups are substantially different from the others. Because these two TEH patterns are relatively straightforward to detect, we reduced sample sizes to create a more challenging setting. Scenarios B and D further introduced within-cluster dispersion: among subgroups belonging to the same true cluster, $\theta_k$ are not identical, and the range is up to 0.16, generating internally less homogeneous clusters. For scenarios with moderate signals, we consider both unbalanced and balanced TEH structures. Scenario E encodes an unbalanced two-cluster configuration where one cluster contains only a few subgroups, while Scenario H represents a balanced partition in which both clusters contain moderate number of subgroups. Scenarios F and H also include within-cluster variance. Additionally, Scenarios G and J incorporate rare boundary subgroups with much smaller sample sizes (20 participants), mimicking real-world situations where certain subpopulations are rare. To evaluate performance when the underlying TEH structure involves three clusters, Scenario K specifies three clusters of a balanced number of subgroups, whereas Scenario L contains two smaller clusters only moderately separable.

\subsection{Simulation Results}
\label{ss:simulation_results}

To compare the performance in estimating subgroup-specific treatment effects, we assess root mean squared error (RMSE) and average interquartile range (IQR).  To compare the ability to detect TEH structure, we evaluate the distribution of $q$ of BHARP and BLAST, and average co-clustering probabilities of BHARP. Additional convergence diagnostics for the rjMCMC algorithm are reported in the Supplementary Material, S2.

%runtime
We implemented the BHARP model via a custom C++ rjMCMC sampler. The simpler comparators (IND, BHM, BLAST) are implemented in Stan, representing their typical usage in practice. 
BART is implemented using the \texttt{BART} package with default settings. 
Under these implementations, the ranking of computational time is consistently BHARP $<$ IND $\approx$ BHM $<$ BART $<$ BLAST. Specifically, on a standard laptop with the same parallel computing setup, analyzing 500 simulated data sets requires at most 2.0 minutes for BHARP across all evaluated scenarios, and IND, BHM and BART require 2.1--4.8 minutes; while for BLAST, the shortest runtime observed is 15.5 minutes, representing approximately an order-of-magnitude increase in computational time. For BHARP, the runtime reflects posterior samples obtained after thinning every third draw, while the comparator models achieve the same number of posterior samples with shorter burn-in periods and no thinning. In other words, BHARP performs approximately three times as many MCMC updates in less wall-clock time, highlighting the efficiency of the custom rjMCMC implementation despite its greater model complexity. 

\begin{figure}[]
\centerline{\includegraphics[width=\textwidth]{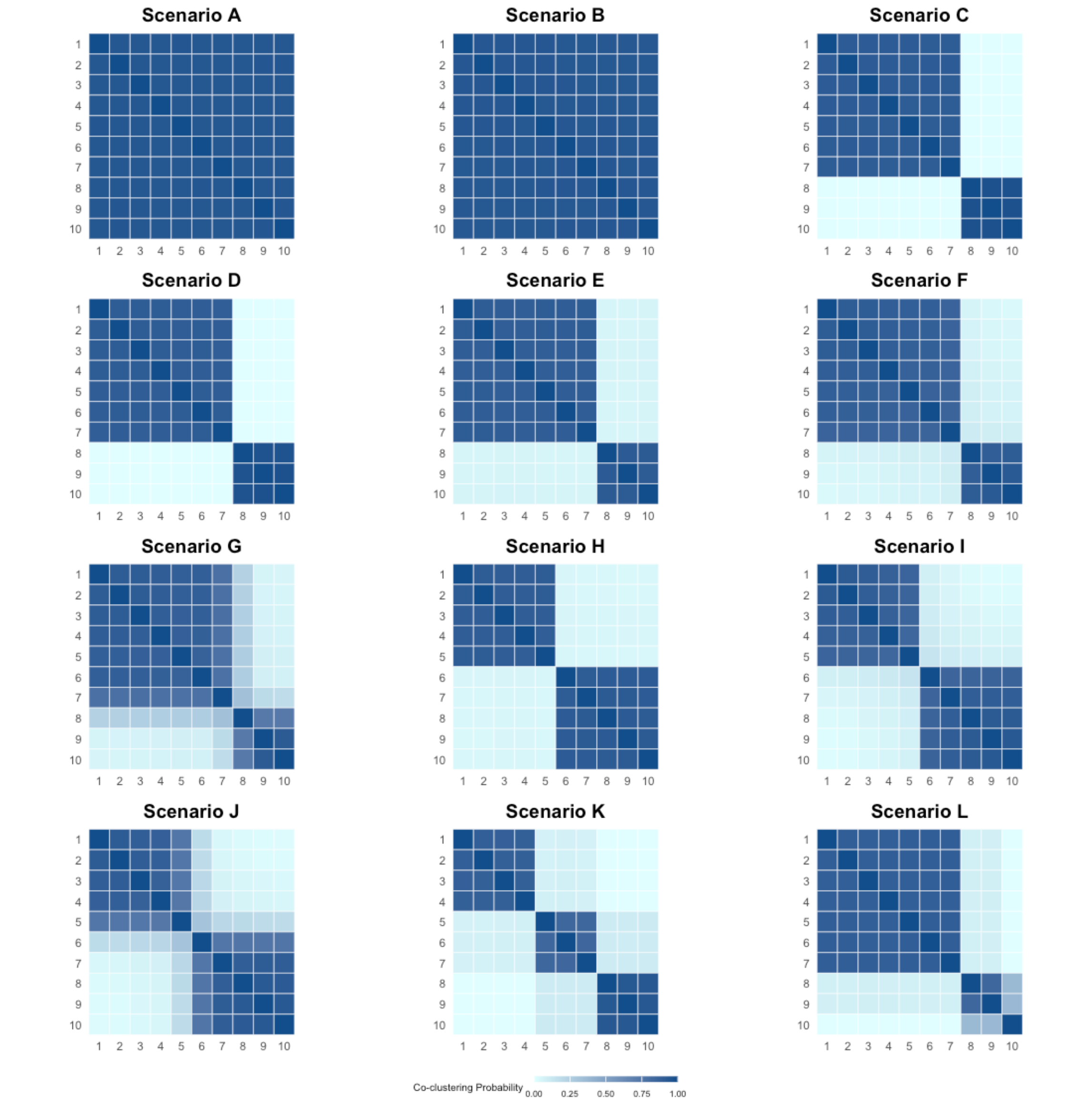}}
\caption{Average co-clustering probability for BHARP. Darker colors indicate stronger evidence that a pair of subgroups belongs to the same component. The true TEH structures are: Scenario A,B \{1,...,10\}; Scenario C--G \{1,...,7\}\{8,9,10\}; Scenario H--J \{1,...,5\}\{6,...,10\}; Scenario K \{1,2,3,4\}\{5,6,7\}\{8,9,10\}; Scenario L\{1,...,7\}\{8,9\}\{10\}. Subgroups 7,8 in Scenario G and 5,6 in Scenario J contain reduced sample sizes. \label{f:BHARPheat}}
\end{figure}

%q and coclustering prob
Among the methods capable of producing partitions---BHARP and BLAST---we compared the empirical distribution of the number of components $q$ and the average co‐clustering probabilities for each subgroup pair. In all scenarios explored, BLAST consistently selected $q=3$ across all MCMC iterations, reducing to a FMM fixed with three components. This behavior limits BLAST’s ability to adapt to the underlying TEH structure, and the resulting components do not necessarily correspond to TEH structure of interest. In contrast, BHARP exhibited adaptive model capacity: the posterior mode of $q$ varies across scenarios, and its empirical distribution (Table~\ref{t:BHARPq}) shows that the most frequently selected value of $q$ closely matches the true number of clusters. Consequently, BHARP provides reliable partition recovery and more flexible levels of information borrowing.
Figure~\ref{f:BHARPheat} shows the average co-clustering probability of each pair of subgroups given by the model which produces sharp block patterns with high within-cluster probabilities and low cross-cluster probabilities, closely recovering the true partition.

\begin{table}[t]
\caption{Distribution of the BHARP posterior mode of the number of components ($q$) across the 500 simulated datasets. Bold entries denote the most frequent posterior mode, which corresponds to the true number of clusters in each scenario.}
\label{t:BHARPq}
\centering
\setlength{\tabcolsep}{4.25pt}
\begin{tabular}{lcccccc}
\toprule
& True & \multicolumn{4}{c}{Frequency of posterior mode} \\
\cmidrule(lr){3-6}
Scenario & clusters & $Mode(q){=}1$ & $Mode(q){=}2$ & $Mode(q){=}3$ & $Mode(q){=}4$ \\
\midrule
A & 1 & \textbf{1.00} & 0.00 & --   & --   \\
B & 1 & \textbf{1.00} & 0.00 & --   & --   \\
C & 2 & --   & \textbf{0.98} & 0.02 & --   \\
D & 2 & --   & \textbf{0.97} & 0.03 & --   \\
E & 2 & 0.01 & \textbf{0.99} & --   & --   \\
F & 2 & 0.01 & \textbf{0.99} & 0.00 & --   \\
G & 2 & 0.01 & \textbf{0.99} & --   & --   \\
H & 2 & 0.00 & \textbf{1.00} & --   & --   \\
I & 2 & --   & \textbf{1.00} & 0.00 & --   \\
J & 2 & --   & \textbf{0.99} & 0.01 & --   \\
K & 3 & --   & 0.02 & \textbf{0.97} & 0.01 \\
L & 3 & --   & 0.35 & \textbf{0.65} & 0.00 \\
\bottomrule
\end{tabular}
\end{table}

\begin{figure}[]
\centerline{\includegraphics[width=\textwidth]{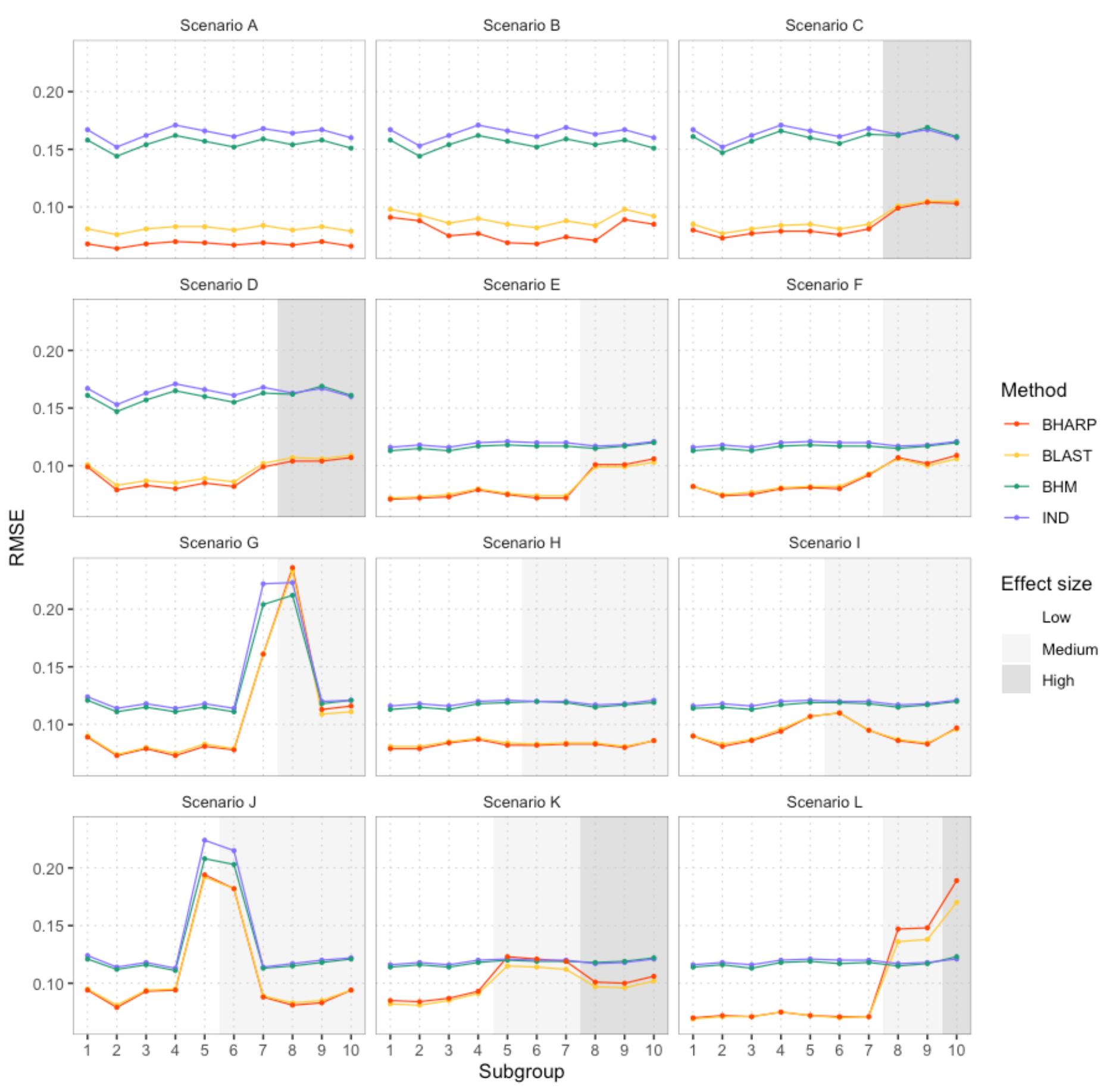}}
\caption{Root mean squared error (RMSE) of posterior median estimates of the BHARP model comparing with the BLAST, BHM, and IND methods. Gray shading indicates true effect sizes of subgroups. Scenarios B,D,F,G,I and J include within-cluster variances. Subgroups 7,8 in Scenario G and 5,6 in Scenario J contain reduced sample sizes. \label{f:RMSE}}
\end{figure}

%RMSE IQR
Figure~\ref{f:RMSE} presents the RMSE of the evaluated methods across the scenarios, with background shading indicating the true effect sizes. The performance of BART is almost indistinguishable from BHM, so it is excluded from graphical summaries to maintain readability. 
Posterior uncertainty summarized by IQR shows consistent patterns: BHARP yields low IQR across all scenarios, indicating more concentrated posterior distribution around subgroup-level responses. Full IQR plots are provided in the Supplementary Material, S3.

%IND BHM BART vs BHARP BLAST
Notably, BHARP and BLAST differ substantially from the other methods, underscoring the benefits of FMM-based partitioning. The performance of the BHM closely resembles that of the IND model, indicating that the between-cluster layer in the BHARP model induces only trivial information borrowing. %BART also performs similarly, as the scenarios include only subgroup indicators and therefore provide limited opportunity for nonparametric methods. Additionally, BART does not directly provide TEH structure without post hoc analysis. 
%BHARP vs BLAST
Overall, when the true number of clusters differs from three, BHARP performs comparably to or slightly better than BLAST. In scenarios where the underlying TEH structure consists of a single homogeneous cluster, BHARP’s flexible prior enables shrinkage towards $q=1$, achieving improved accuracy and precision relative to BLAST, whose three‐component formulation induces unnecessary splitting. Conversely, when the true TEH structure matches BLAST’s built-in specification of three clusters, BHARP does not fall behind: its RMSE remains nearly indistinguishable across the majority of subgroups, demonstrating that the added flexibility of model capacity does not compromise performance even when the component number in fixed FMM is correctly specified. 

BHARP and BLAST accurately recover the underlying TEH structure in scenarios where subgroups are either highly homogeneous or exhibit pronounced between-cluster differences, even when subgroup sample sizes are limited (Scenario A--D). 
In challenging settings where ``border subgroups" contain smaller sample sizes and the transition between clusters becomes less distinct, RMSE increases for all methods (Scenario G and J). In scenarios with balanced cluster sizes (Scenarios H--J), BHARP and BLAST nevertheless maintain clear RMSE advantages. However, when signals are moderate and the true clusters are unbalanced (Scenarios E--G), both methods experience increased difficulty in delineating the smaller cluster, especially when ``border subgroups" are sparsely sampled. Heatmaps of co-clustering probabilities (Figure~\ref{f:BHARPheat}) corroborate this pattern, showing diffuse boundaries in Scenario G. A similar phenomenon appears in three-cluster scenarios: when two adjacent clusters have modest separations and contain substantially fewer subgroups than the dominant cluster, both BHARP and BLAST display some difficulty recovering precise boundaries, although they still outperform non-FMM approaches in overall performance (Scenario L).

%BLAST
The original BLAST framework is a joint model for binary response and longitudinal biomarker measurements, equipped with latent-class–based information borrowing and cluster-specific penalized-spline trajectories. 
Although we refer to our comparator as ``BLAST", our implementation retains only the FMM-based information borrowing and the DIC-based model selection procedure. In this simplified setting, DIC becomes even more prone to favor larger values of $q$ than in the original formulation. This is a direct consequence of the model–criterion combination: increasing $q$ does not increase likelihood parameters, so the effective complexity penalty changes little. As a result, DIC deterministically selects the maximal candidate $q=3$.

Additionally, our comparative study concerns the generalizability of the model-selection strategy adopted in BLAST framework. 
First, the original BLAST framework restricts the candidate values of $q$ to $\{1,2,3\}$. This restriction reflects a practical limitation: without such limitations, manually fitting and comparing a large collection of models would become unwieldy. 
Second, even within this restricted search space, the selection procedure is computationally slow. For a single-arm analysis, fitting three candidate models already requires noticeably more computation time. The burden is even exacerbated when extending BLAST's model-selection strategy to multi-arm or platform trials: each arm requires a separate search, so the total number of candidate models grows exponentially (e.g., $3^I$ candidate models for $I$ intervention arms). In contrast, BHARP automatically learns the borrowing structure together with parameter value without reliance on manual selection. The computational cost scales linearly with the complexity of trial design, making BHARP particularly efficient for large platform trials.
Third, BLAST relies on model selection statistics, which have become increasingly controversial. Furthermore, DIC implicitly assumes the posterior distribution to be approximately unimodal and symmetric, which is likely to be violated in FMMs. Such violations can lead to unstable or biased model comparisons and a tendency to favor over-parameterized models.
In conclusion, the original BLAST paper explicitly noted rjMCMC as an alternative approach to select $q$ but did not implement it. BHARP fills this gap by incorporating rjMCMC into a dynamic partitioning scheme, allowing the uncertainty in cluster configuration and information borrowing to be integrated into posterior inference.

\section{Application to the Partner Step T2D Clinical Trial}
\label{s:application}

We illustrate the application of BHARP framework using a design inspired by the ``Partner Step T2D" trial, which evaluates behavioral interventions to increase the step counts among adults with type 2 diabetes (T2D) and their partners. In the actual trial, couples receive wearable step counters and either weekly step targets alone or targets plus a couple-based behavioral coaching (``dyadic coping") intervention. 
While the actual trial involves two intervention arms, for illustrative purposes we consider a three-arm design to showcase BHARP framework’s ability to capture distinct TEH patterns. We simulate an adaptive enrichment design by adding an arm with step counters only. TEH patterns are assumed to depend on two baseline factors: relationship quality (low/medium/high) and concordance for excess weight (no/yes, defined as both partners having body mass index $\geq$30 kg/$\text{m}^2$). Crossing these factors yields six subgroups that represent distinct patient profiles potentially influencing the treatment effects of three investigated interventions. Table~\ref{t:applicationDEF} summarizes the definition as well as the true arm-by-subgroup means used for data generation are summarized, which encodes homogeneous weak effects (Arm 1), gradient effects aligned with marital quality (Arm 2), and stronger effects among concordant couples (Arm 3).

\begin{table}[t]
\caption{Subgroup definitions and true arm-by-subgroup means $(\theta_{ik})$ used in the application data-generating process. Subgroups are defined by crossing marital quality (low/medium/high) and concordance for excess weight (no/yes; both partners having BMI $\ge 30$ kg/$\text{m}^2$). Outcomes are generated as $Y_{ik} \sim \mathcal{N}(\theta_{ik},1)$.}
\label{t:applicationDEF}
\centering
\setlength{\tabcolsep}{4.5pt}
\begin{tabular}{lcccccc}
\toprule
Subgroup & 1 & 2 & 3 & 4 & 5 & 6 \\
\midrule
Marital Quality & Low & Medium & High & Low & Medium & High \\
Concordance for Excess Weight & No & No & No & Yes & Yes & Yes \\
\midrule
True means $(\theta_{ik})$, Arm 1 & 0.10 & 0.10 & 0.10 & 0.10 & 0.10 & 0.10 \\
True means $(\theta_{ik})$, Arm 2 & 0.10 & 0.40 & 0.70 & 0.13 & 0.43 & 0.73 \\
True means $(\theta_{ik})$, Arm 3 & 0.60 & 0.65 & 0.70 & 1.20 & 1.22 & 1.25 \\
\bottomrule
\end{tabular}
\end{table}

We consider a hypothetical adaptive enrichment design with interim analyses at cumulative sample sizes of 1200, 1500, and 1800, followed by a final analysis at 2100. The futility and efficacy boundaries are $x_F=0.15, x_E=0.5$, and the posterior probability thresholds are $P_F=0.5, P_E=0.95$, respectively. 
The futility boundary $x_F$ represents a minimal effect below which continued accrual is not justified, while the efficacy boundary $x_E$ represents a more stringent compelling effect to declare success. Using $x_E>x_F$ makes efficacy conclusions quite conservative and typically requires larger sample sizes. 
Decisions are made based on the analysis results of BHARP, BHM, and IND models. Evaluation metrics include the global false positive rate, generalized power, expected total sample size, RMSE, and IQR.
Generalized power is defined as the probability that at least one effective subgroup is correctly identified for each arm, or that at least one effective intervention is identified for each subgroup. 
Hyperparameter settings follow Section~\ref{ss:hyperparam}, except for the within-component variance, for which we use a more informative prior ($a_{\text{within}}=150$, $b_{\text{within}}=1.51$) to accommodate the larger sample size.

\begin{figure}[]
\centering
\includegraphics[width=\textwidth]{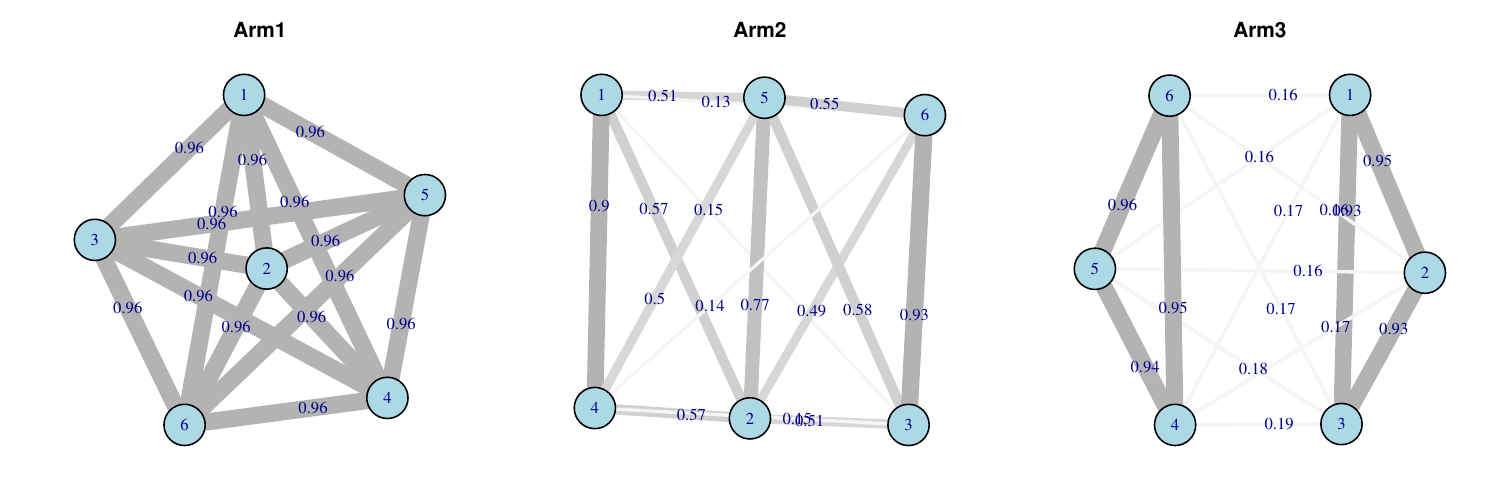}
\caption{Network plots of co-clustering probabilities estimated by BHARP model.
Thickness and color of edges represent the average co-clustering probability between subgroup pairs.}
\label{f:network}
\end{figure}

We first illustrate the utilities of the BHARP framework for detecting TEH patterns using the average results across 500 simulated trials.
Figure~\ref{f:network} displays the pairwise co-clustering probabilities, where the width and color of edges indicate the average posterior probability that two subgroups are assigned to the same mixture component. BHARP successfully recovers the underlying TEH structure in all three arms: a single homogeneous cluster in Arm 1, three cross-paired clusters in Arm 2 ($\{1,4\}, \{2,5\}, \{3,6\}$) and two larger clusters in Arm 3 ($\{1,2,3\}, \{4,5,6\}$). Although distinguishing weakly separated small clusters in Arm 2 remains challenging, the estimated co-clustering patterns closely align with the true partitions. In addition, BHARP requires less than half the computational time compared with BHM and IND, further highlighting its practical advantage for large-scale adaptive trial simulations.

% network plot 

\begin{figure}[]
\centering
\includegraphics[width=\textwidth]{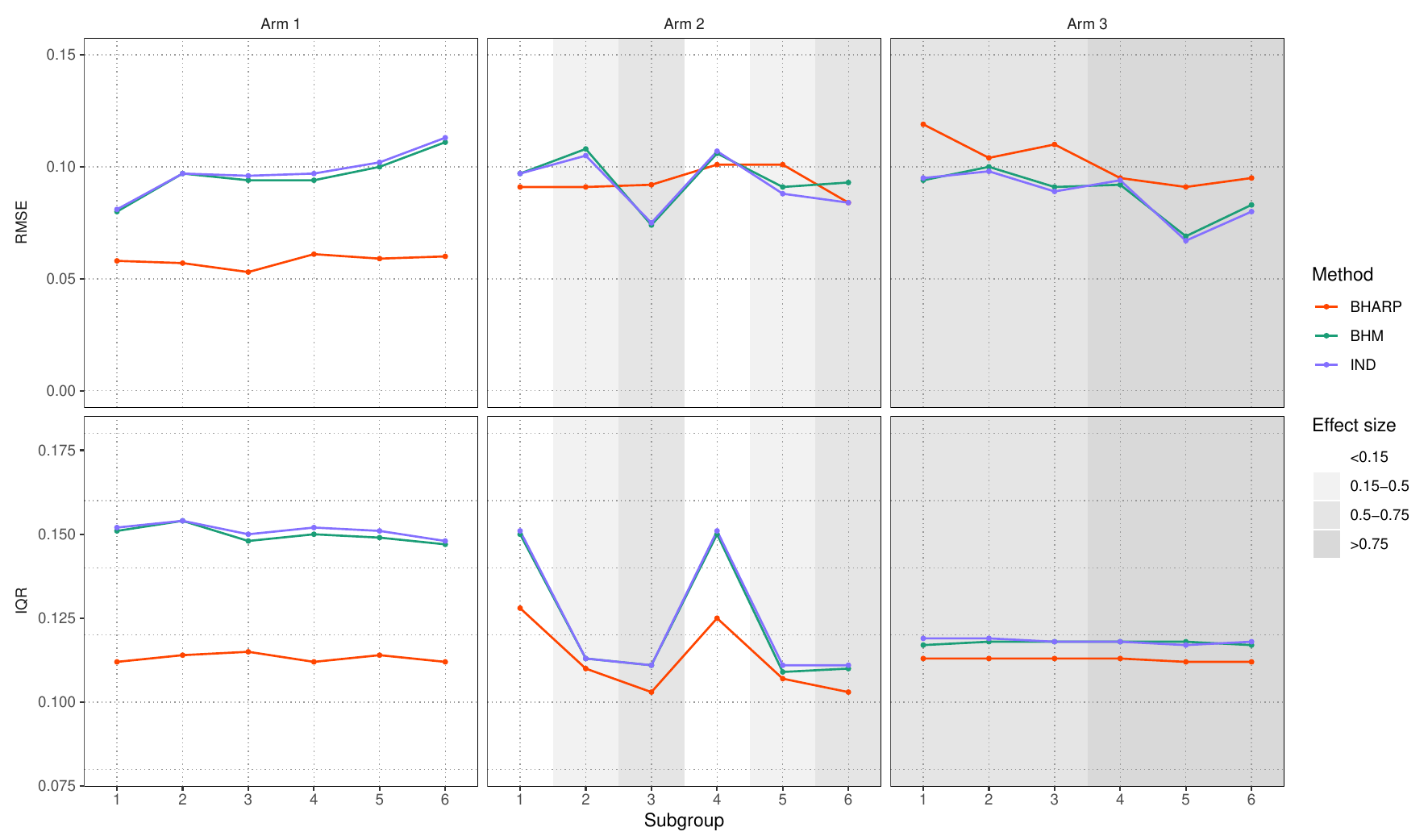}
\caption{RMSE and IQR summarizing the estimation performance of BHARP, BHM, and IND models.
Shaded regions indicate the underlying true effect-size levels.}
\label{f:RMSEtrial}
\end{figure}

Table~\ref{t:applicationRESULTS} compares the operating characteristics under the adaptive enrichment design. 
All three methods control the global false positive rate at approximately 0.02, and in terms of identifying effective subpopulation at the arm level, the three methods exhibit comparable performance. 
However, the improvement offered by BHARP in identifying effective interventions for each subpopulation is substantial. Its generalized power at the subgroup level reaches 0.44—more than twice that of BHM and IND. This indicate that BHARP provides a more granular characterization of heterogeneous subpopulations.  
BHARP also makes faster decisions in Arm 1 (all futile) and Arm 3 (all effective), allowing it to reallocate more samples to Arm 2, where clusters are smaller and futile and effective responses are mixed.
Figure~\ref{f:RMSEtrial} summarizes the estimation performance across methods.
In the absence of TEH in Arm 1, BHARP achieves substantially lower RMSE than the comparator methods. In Arm 3, the RMSE of BHARP is slightly higher than that of BHM and IND, but the differences are small. Given that BHARP accurately recovers the underlying TEH structure in this arm, the modest increase in RMSE is likely attributable to the relatively smaller effective sample size resulting from early stopping.
In addition, although BHM and IND allocate larger sample sizes to Arms 1 and 3, their IQR remain higher, indicating greater posterior uncertainty even with additional data. In contrast, BHARP consistently achieves the lowest IQR across all subgroups in all arms.

\begin{table}[t]
\caption{Comparison of operating characteristics across BHARP, BHM, and IND.}
\label{t:applicationRESULTS}
\centering
\setlength{\tabcolsep}{4.5pt}
\begin{tabular}{lccc}
\toprule
 & BHARP & BHM & IND \\
\midrule
Global false positive rate & 0.02 & 0.02 & 0.02 \\
Generalized power (arm level) & 0.94 & 0.98 & 1.00 \\
Generalized power (subgroup level) & 0.44 & 0.18 & 0.20 \\
Expected sample size (Arm 1) & 478 & 515 & 519 \\
Expected sample size (Arm 2) & 892 & 773 & 770 \\
Expected sample size (Arm 3) & 730 & 810 & 810 \\
\bottomrule
\end{tabular}

\vspace{2mm}
\footnotesize
Generalized power (arm level) refers to the probability that each arm has at least one truly effective subgroup correctly identified.
Generalized power (subgroup level) refers to the probability that each subgroup has at least one truly effective intervention correctly identified.
\end{table}

%application RMSE 

\section{Discussion}
\label{s:discussion}

%proposals
In this paper, we proposed BHARP, a Bayesian framework that identifies TEH structures while adaptively borrowing information across subgroups. The model jointly explores clustering configurations and subgroup-level effects, yielding coherent posterior summaries of TEH and offering a practical implementation of rjMCMC suitable for adaptive enrichment trials.
BHARP partitions subgroups using a finite mixture model with an unknown number of components, allowing the data to determine the structure of information borrowing. In simulations, BHARP achieved low RMSE and IQR for subgroup-specific estimates,  offered greater flexibility than FMM with a fixed number of components without sacrificing performance, and its computational efficiency exceeded that of Bayesian hierarchical and independent models.
In the multi-arm adaptive trial application, the BHARP model recovered the underlying TEH structure and demonstrated desirable operating characteristics, including appropriate false positive control, the ability to identify effective interventions for subpopulations, and timely arm termination.

% practical considerations
From a practical standpoint, several considerations are important when incorporating TEH into trial design using BHARP. The interpretability of TEH depends critically on the prespecified subgroup structure: overly coarse groupings may mask meaningful heterogeneity, whereas excessively fine groupings can dilute the available information within each subgroup. In addition, when TEH signals are expected to be weak, reliable detection may require larger sample sizes or carefully elicited informative priors to ensure stable estimation. These considerations highlight that, while BHARP provides a flexible framework for TEH modeling, its performance and interpretability remain intrinsically tied to the quality of subgroup definition and the amount of available information.
% From a practical perspective, BHARP also highlights several considerations for incorporating TEH into clinical trial design. First, the interpretability of TEH depends critically on the prespecified subgroup structure. BHARP can only reveal heterogeneity across predefined subpopulations, making clinically informed subgroup definitions essential for meaningful inference.
% Second, detecting subtle TEH requires sufficient information. When heterogeneity signals are expected to be weak, larger sample sizes or carefully elicited informative priors become important for achieving stable estimation.
% Finally, the performance of BHARP depends on the granularity of the prespecified subgroups. Very coarse subgroups may mask meaningful heterogeneity, whereas excessively fine subgroups can dilute the available information within each subgroup. Selecting a granularity that reflects clinically relevant variation while preserving adequate information within each subgroup is therefore essential for reliable TEH identification.

%advantage:TEH
A key advantage of BHARP is its explicit integration of model uncertainty. By jointly sampling model dimensionality and subgroup allocations, BHARP avoids conditioning inference on a single clustering structure. The posterior samples also enable intuitive summaries of TEH—such as pairwise co-clustering probabilities—and more targeted queries regarding clinically defined subpopulations. This is particularly valuable in precision medicine, where understanding the structure of heterogeneity is often as important as estimating subgroup-level effects.
%A notable advantage of the BHARP framework is its principled integration of model uncertainty. Unlike conventional models, BHARP jointly samples model dimensionality and subgroup allocations through reversible-jump transitions. This enables posterior inference to reflect uncertainty in the TEH configuration rather than conditioning on a pre-specified clustering structure. Beyond estimating subgroup-level responses, the sampling-based characterization of the partition space enables empirical evaluation of TEH features. While we presented pairwise co-clustering probabilities as an intuitive summary, BHARP naturally supports richer queries, such as the posterior probability that a clinically defined subset forms a coherent cluster that is distinct from other subgroups. Such capabilities are particularly valuable in precision medicine, where identifying the structure of heterogeneity is as important as estimating subgroup-specific effects. 

%advantage:DPMM
Compared to Dirichlet process mixture models (DPMMs) with potentially unbounded number of components, BHARP provides a more transparent balance between flexibility and interpretability. Finite mixture models with an unknown number of components share many theoretical advantages of DPMMs \citep{MillerHarrison2018}, but avoid reliance on generative schemes such as the Chinese restaurant process, which can complicate prior elicitation. BHARP’s hyperparameters offer more direct control over borrowing strength and can incorporate clinical judgment on interpretable scales.
%In addition, BHARP offers a practical advantage over Dirichlet process mixture models (DPMMs), which assume a potentially unbounded number of components. Many of the desirable theoretical properties of DPMMs are also shared by finite mixtures with an unknown number of components (\citealp{MillerHarrison2018,geng2020}). However, the mechanism through which DPMMs generate new clusters, such as the Chinese restaurant process, can be challenging for eliciting or incorporating prior clinical knowledge. In contrast, BHARP provides a more transparent balance between flexibility and practical control, allowing users to incorporate expert judgment through hyperparameters with interpretable scales.

%advantage: greater I and K
BHARP is also well suited for modern large-scale settings such as multi-arm or platform trials. Our implementation provides an efficient R interface requiring no manual specification of the borrowing structure. Unlike approaches that first perform clustering and then condition on a selected partition, BHARP integrates partition uncertainty directly into posterior inference, yielding a fully automated workflow that becomes increasingly valuable as the number of treatment arms and subgroups grows. In practice, increasing the number of regimens does not lead to exponential growth in computation time, while the expanded partition space makes enumeration or selection of a single optimal partition impractical and necessitates posterior sampling. Furthermore, richer subgroup-level information reduces uncertainty in the partition structure and thereby improves BHARP’s ability to recover TEH patterns.
%The rjMCMC algorithm is computationally efficient, and increasing the number of treatment arms does not lead to an exponential growth in computation time. Our implementation provides a simple R interface that requires no manual specification of the information-borrowing structure. Unlike approaches that first fit a clustering model and then condition subsequent borrowing on a selected partition, BHARP integrates partition uncertainty directly into posterior inference, yielding a fully automated workflow. This automation becomes increasingly valuable as the number of subgroups grows, expanding the partition space to the point where enumeration or selecting a single optimal partition becomes impractical. In fact, the presence of richer subgroup-level information further enhances BHARP’s ability to characterize TEH. As a result, the method is particularly well aligned with modern multi-arm, multi-domain platform trials, where both the number of treatment regimens and the granularity of patient stratification continue to expand.

%computational contribution: 
A substantial part of this work concerns the computational development required to make rjMCMC viable for dynamic partitioning in clinical-trial settings.  Although rjMCMC is well established in areas such as variable selection, its application to partition-based information borrowing has remained limited. Clinical trials often involve modest effect sizes and limited sample sizes, conditions under which poorly constructed transdimensional proposals can lead to unstable or clinically implausible partitions. To address these challenges, we adapted the split–merge framework \citep{nobilegreen2000} with modifications guided by the typical sample sizes encountered in clinical trials. Because these modifications target general features of clinical trials, they are expected to generalize beyond the specific settings considered here. These enhancements stabilize exploration of the partition space and improve the algorithm’s ability to detect subtle heterogeneity, thereby making rjMCMC a practical tool for TEH modeling in applied trial settings.
%A substantial part of BHARP’s contribution lies in the computational development required to make rjMCMC viable for dynamic partitioning in clinical trial settings. Although rjMCMC is well established for variable selection, its application to partition-based information borrowing has remained limited. Clinical trials typically involve modest effect sizes and limited sample sizes, conditions under which poorly constructed transdimensional proposals can easily produce unstable, incoherent, or clinically implausible partitions. Designing reversible split–merge moves that remain both statistically valid and practically credible is therefore nontrivial. To address these challenges, we adapted the classical split–merge framework \citep{nobilegreen2000} with several domain-specific modifications, including an informative prior to encourage conservative borrowing, and an increased frequency of split proposals to mitigate over-merging. These enhancements allow the algorithm to explore clinically meaningful partitions and increase its ability to detect subtle heterogeneity.

%limitations and future work
The method can be extended in several directions. 
One avenue is to define subgroups in a data-driven manner. Approaches such as Bayesian model averaging for biomarker selection and threshold discovery \citep{maleyeff2024} could be integrated into BHARP to support data-driven subgroup formation. 
A second direction concerns longitudinal outcomes, where repeated measurements can be viewed as a large number of small ``subgroups" in the BHARP framework. Extending BHARP to this formulation would enable the identification of subject-level TEH across individual response trajectories. Related ideas have been explored in joint models such as BLAST \citep{ChuYuan2018blast} and represent a challenging but important next step.
Finally, while BHARP provides flexible inference on TEH structures, these posterior summaries are not yet incorporated into adaptive decision rules. A promising extension is to allow adaptive allocation among active interventions rather than using equal allocation, so that additional enrollment is focused on subgroups or interventions where TEH patterns remain poorly resolved.

\section*{Supplementary Materials}
Supplementary Materials are available in the ancillary files of this arXiv submission.

\section*{Acknowledgements}
This research was supported by a doctoral scholarship from the Fonds de recherche du Québec – Nature et technologies (FRQNT). SG acknowledges support from the Natural Sciences and Engineering Research Council of Canada (NSERC), Canadian Institute for Statistical Sciences (CANSSI), Fonds de recherche du Québec - Santé (FRQS) and FRQNT. 
The Partner Step T2D trial was supported by Diabetes Canada.  The authors acknowledge the use of computing resources provided by the Digital Research Alliance of Canada.

\vspace*{-8pt}

\section*{Data availability statement}
The code for the proposed method and scripts to reproduce the simulation studies 
will be available at \href{https://github.com/Xianglin-XLZ/BHARP}{GitHub repository}. Simulation code is available upon request and will be made public upon acceptance.
\vspace*{-8pt}

\bibliographystyle{unsrtnat}
\bibliography{biosample}

\end{document}